\RequirePackage{rotating}
\documentclass[utf8,a4paper]{frontiersFPHY} 

\setcitestyle{square} 
\usepackage{url,hyperref,microtype}
\usepackage[onehalfspacing]{setspace}
\usepackage{units}

\usepackage{graphicx}
\usepackage{epstopdf}
\usepackage{amsmath}

\def\keyFont{\fontsize{8}{11}\helveticabold }
\def\firstAuthorLast{Kiefer {et~al.}} 
\def\Authors{Ren\'{e} Kiefer\,$^{1,*}$, Anne-Marie Broomhall\,$^{1}$, Warrick H. Ball\,$^{2,3}$}
\def\Address{$^{1}$Centre for Fusion, Space, and Astrophysics, Department of Physics, University of Warwick, Coventry, United Kingdom\\
	$^{2}$School of Physics and Astronomy, University of Birmingham, Birmingham, United Kingdom\\
	$^{3}$Stellar Astrophysics Centre, Department of Physics and Astronomy, Aarhus University, Aarhus, Denmark}
\def\corrAuthor{Ren\'{e} Kiefer}
\def\corrEmail{r.kiefer@warwick.ac.uk}

\begin{document}
	\onecolumn
	\firstpage{1}
	
	\title[Seismic Signatures of Stellar Magnetic Activity]{Seismic Signatures of Stellar Magnetic Activity --- What Can We Expect from TESS?} 
	
	\author[\firstAuthorLast]{\Authors} 
	\address{} 
	\correspondance{} 
	
	\extraAuth{}
	
	\maketitle
	
	\begin{abstract}
		Asteroseismic methods offer a means to investigate stellar activity and activity cycles as well as to identify those properties of stars which are crucial for the operation of stellar dynamos. With data from CoRoT and \textit{Kepler}, signatures of magnetic activity have been found in the seismic properties of a few dozen stars. Now, NASA’s Transiting Exoplanet Survey Satellite (TESS) mission offers the possibility to expand this, so far, rather exclusive group of stars. This promises to deliver new insight into the parameters that govern stellar magnetic activity as a function of stellar mass, age, and rotation rate. We derive a new scaling relation for the amplitude of the activity-related acoustic (p-mode) frequency shifts that can be expected over a full stellar cycle. Building on a catalogue of synthetic TESS time series, we use the shifts obtained from this relation and simulate the yield of detectable frequency shifts in an extended TESS mission. We find that, according to our scaling relation, we can expect to find significant p-mode frequency shifts for a couple hundred main-sequence and early subgiant stars and for a few thousand late subgiant and low-luminosity red giant stars.
		
		\tiny
		\keyFont{ \section{Keywords:} stellar activity, asteroseismology, stellar activity cycles, TESS, p-mode frequency shifts} %
	\end{abstract}
	
	\section{Introduction}\label{sec:intro}
	The primary seismic signature of stellar activity and activity cycles is a systematic change in mode frequencies over timescales associated with activity cycles, which are typically of the order of months to decades \cite{1995ApJ...438..269B, 1999ApJ...524..295S, 2014MNRAS.441.2744V}. Measuring this phenomenon for stars is primarily limited by the baseline length of the observations. For the Sun, however, we have Sun-as-a-star helioseismic observations lasting for decades, covering several eleven-year solar activity cycles. From these data it is found that the frequency of low-degree modes close to the frequency of maximum oscillation power $\nu_\textrm{max}$ change by approximately $\unit[0.4]{\mu Hz}$ between activity minimum and maximum (e.g., \cite{1998A&A...329.1119J,2017SoPh..292...67B}). Furthermore, this shift is strongly correlated with the level of activity along the cycle \cite{1998A&A...329.1119J, 2015SoPh..290.3095B}. 
	
	The first asteroseismic evidence of variations in stellar activity were found in an F5\,V star observed by CoRoT  \citep{2006ESASP1306...33B, 2009A&A...506..411A}, HD\,49933. For this star \citet{2010Sci...329.1032G} found that not only did the frequencies vary systematically with time but, as is also observed for the Sun \cite{palle90, 2000ApJ...543..472K, 2015SoPh..290.3095B}, the amplitudes of the modes varied in anti-phase with the frequencies. In a more detailed analysis, \citet{2011A&A...530A.127S} found that the observed frequency shifts are larger than seen in the Sun but the same dependence on mode frequency is observed, i.e., the magnitude of the frequency shift increases with mode frequency (e.g., \cite{1990Natur.345..779L, 2004A&A...413.1135S}).
	
	Since then examples of systematic variations in the frequencies of stars observed by \textit{Kepler} \citep{Borucki977, 2010ApJ...713L..79K} have been found (\citet{2016A&A...589A.118S, 2016A&A...589A.103R, 2017A&A...598A..77K, 2018A&A...611A..84S, 2018ApJS..237...17S}). Indeed, \citet{2017A&A...598A..77K} found that 23 out of the 24 stars in their sample showed significant and systematic frequency shifts in time, while \citet{2018ApJS..237...17S} found quasi-periodic variations on 60\% of their sample of 87 solar-type stars. The sample of potential candidates for observing asteroseismic activity cycles is limited by the fact that activity suppresses the amplitude of acoustic oscillations \cite{2011ApJ...732L...5C}. Since activity also increases with decreasing rotation period \cite[e.g.,][]{2007ApJ...657..486B, 2008ApJ...687.1264M, 2016A&A...595A..12S, 2017SoPh..292..126M} these stars tend to be more active and consequently are less likely to have high signal-to-noise oscillations. The dependence of the length of stellar activity cycles on surface rotation rate or on Rossby number is complex and is actively researched \cite[e.g.,][]{2017Sci...357..185S, 2018A&A...616A..72W, 2018ApJ...863...35S}. While it has proven to be difficult to observe complete activity cycles with \textit{Kepler}, it is still possible to search for systematic variations with time in the properties of p modes. A concise review on the capabilities and prospects of asteroseismology regarding the inference on stellar cycles and activity is also given by \citet{2014SSRv..186..437C}.
	
	Stellar magnetic activity also manifests in variability of stellar luminosity \cite[see, e.g.,][and references therein]{2017AN....338..753F}. Over the solar cycle the total irradiance is found to change by about 0.1\% in temporal correlation with solar activity  \cite{2013SSRv..176..237F}. This slight increase in luminosity is associated with faculae \cite{2013SSRv..176..237F}. In contrast to faculae, spots decrease solar and stellar luminosities \cite{2013SSRv..176..237F, 2017NatAs...1..612S}. For the Sun, the increase in luminosity caused by faculae dominates over the decrease in luminosity brought about by sunspots, hence the overall increase in total irradiance between cycle minimum and maximum. However, for other stars it is the decrease in luminosity caused by the presence of spots that is the dominant source of variability \citep{2014A&A...569A..38S}. Magnetic features have typical life times which are of the order of the rotation period of a star. This has to be taken into account when a meaningful proxy of stellar magnetic activity is to be constructed from the variability of the stellar luminosity. Indeed, \citet{2014JSWSC...4A..15M} defined such a photometric proxy for stellar activity with the $S_{\text{ph}}$ index. It is given by taking the mean of the standard deviations of segments of the time series each having a length 5 times the rotation period of the star (see also \cite{2016A&A...596A..31S} and \cite{2017A&A...608A..87S}). As long cadence photometric time series (CoRoT: \unit[512]{s}, \textit{Kepler}: \unit[29.4]{min}, TESS: \unit[30]{min}) are sufficient for such studies, the number of available light curves for the investigation of stellar magnetic activity through photometry is rather large (CoRoT: $\sim$163\,000 targets, \textit{Kepler}: $\sim$197\,000 targets, TESS: $>10^7$ targets at end of nominal mission). Studies in which the variability of light curves is explored regarding signatures of activity or activity cycles include, but are not limited to \cite{2014MNRAS.441.2744V,2015A&A...583A.134F,2016A&A...596A..31S,2017ApJ...851..116M,2019MNRAS.485.5096K}. In this article, we do not consider the variability of light curves, as activity-related changes are not included in the synthetic time series we use (see \cite{2018ApJS..239...34B} and Section~\ref{ssec:TESS_target}).
	
	With the launch of NASA’s Transiting Exoplanet Survey Satellite \cite[TESS,][]{2015JATIS...1a4003R}, the opportunity of studying seismic signatures of many more stars is at hand. So far, p-mode frequency shifts have only been detected for a few dozen \textit{Kepler} targets and one CoRoT target, and as of yet, the dependence of the amplitude of the activity-related frequency shifts (which corresponds to the amplitude of stellar activity cycles) on fundamental stellar parameters is not known. With the large number of stars that will be observed by TESS, asteroseismology can become an important additional tool to learn about stellar cycles. As we will discuss in Section~\ref{sec:TESS}, a mission extension of TESS to four or six years is however necessary for this. This paper is aimed at a prediction of how many detections of activity-related frequency shifts we can expect in an extended TESS mission.  
	
	\section{Synthetic TESS sample}\label{ssec:TESS_target}
	To simulate the yield of seismic signatures of activity that can be expected from TESS, we use the synthetic catalogue of \citet{2018ApJS..239...34B}, which consists of light curves that realistically mimic the properties of the TESS targets in short-cadence data (\unit[120]{s}). The sample consists of 12\,731 stars, covering main-sequence, subgiant, and low-luminosity red giant stars, which were selected from a synthetic Milky Way population simulated with TRILEGAL \citep{2005A&A...436..895G}. The selection criteria are the same as for the real Asteroseismic Target List (ATL; see \citet{2019ApJS..241...12S}) of the TESS Asteroseismic Science Consortium (TASC), which are chosen such as to maximise the yield of detected oscillations and optimise the coverage of different stages of stellar evolution.  \citet{2018ApJS..239...34B} give a detailed description of their methods to obtain oscillation mode parameters (frequencies, mode damping widths, amplitudes), noise levels, granulation background, and other aspects of the simulated time series. We note that the time series of \citet{2018ApJS..239...34B} are provided without noise, which has to be added as $\mathcal{N}\left(0,\sigma_{\text{noise}}\right)\cdot\sqrt{30}$ to the time series. Here, $\sigma_{\text{noise}}$ is the noise amplitude as given in the catalogue of  \citet{2018ApJS..239...34B} and the factor $\sqrt{30}$ is accounting for the 2-minute sampling of the time series.
	
\begin{figure}
		\begin{center}
			\includegraphics[width=0.8\textwidth]{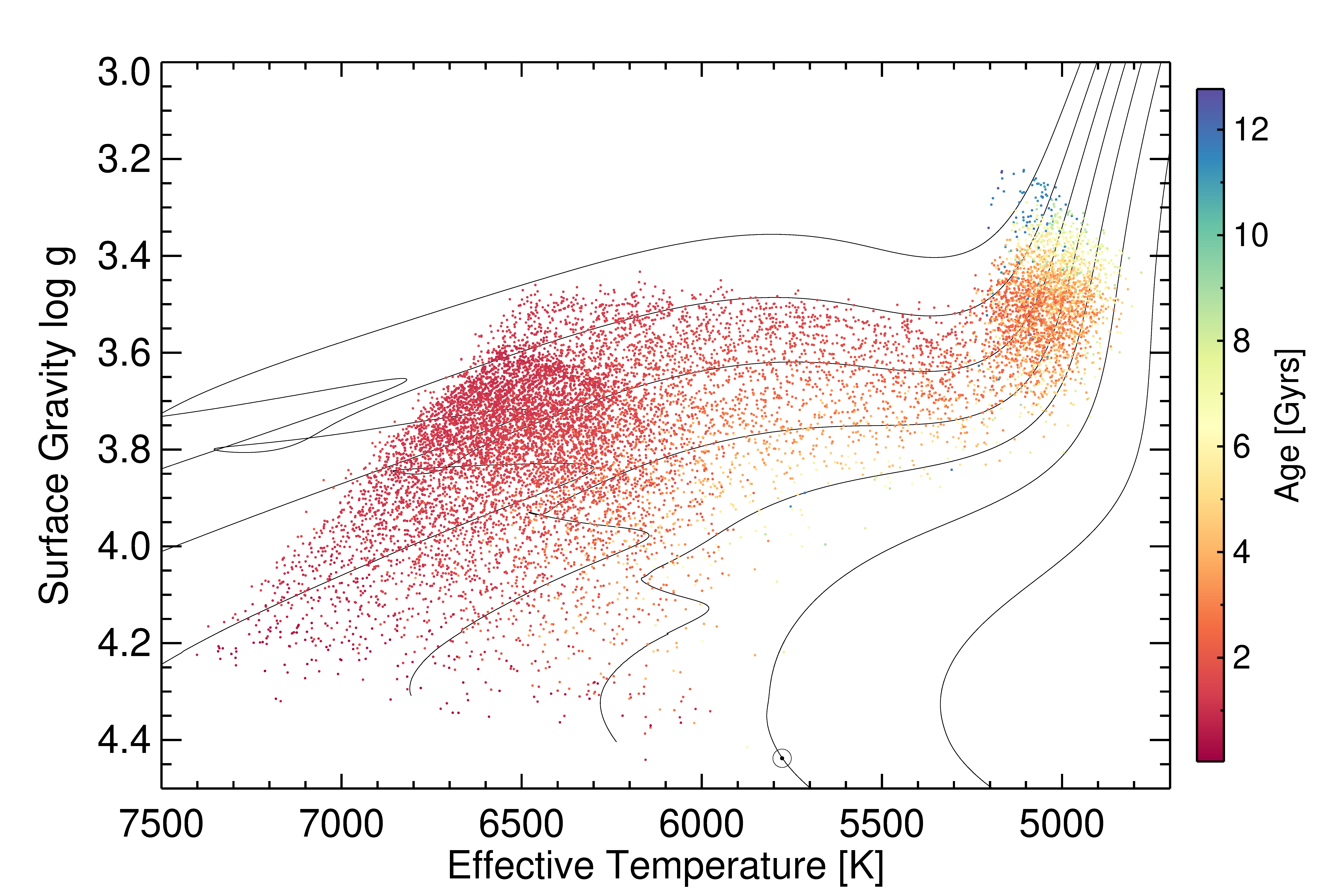}
			\caption{Kiel diagram of the synthetic TESS sample of \citet{2018ApJS..239...34B}. The age of the stars is given by the colour of the dots. The Sun is indicated by its usual symbol. Evolutionary tracks for stars with masses from 0.8 to \unit[2.0]{M$_{\odot}$} in steps of \unit[0.2]{M$_{\odot}$} are overlaid as solid black lines.}
			\label{fig:MATL_Kiel}
		\end{center}
	\end{figure}

	Figure~\ref{fig:MATL_Kiel} shows a Kiel diagram of the synthetic sample with stellar ages given by the colour of the dots. We separated the sample into main-sequence stars (in the following called region I) and post-main-sequence stars (region II) along the Terminal Age Main-Sequence (TAMS), which we defined at a core hydrogen abundance content of $10^{-5}$. Figure~\ref{fig:MATL_Kiel_TAMS} shows the Kiel diagram as in Figure~\ref{fig:MATL_Kiel} but colour coded for the two regions. As can be seen in these figures, the number of stars on the lower main-sequence is rather small. This is due to the low intrinsic mode amplitudes of solar-like oscillations in cool dwarf stars (see, e.g., \citet{1995A&A...293...87K}). Thus, these stars have a low detection probability for their oscillation modes with TESS data \citep{2016ApJ...830..138C}. A large number of targets are young, early subgiants and stars towards the end of their main-sequence life with $\unit[6000]{K}\lesssim T_{\text{eff}}\lesssim \unit[6900]{K}$. Subgiants ascending to the red giant branch and low-luminosity red giants have large mode amplitudes and high detection probabilities, hence, many of the best asteroseismic targets are in this region of the Kiel diagram.
	
	For a study of the temporal evolution of seismic parameters, it is necessary that the targets are observed for either a long, uninterrupted period of time or that they are observed again after some time. During its nominal two year mission, TESS will observe 13 sectors per hemisphere with an area of four times $24^{\circ}\times24^{\circ}$ for a duration of about 27.4 days per sector. The sectors partly overlap, chiefly in the region of the ecliptic poles. Thus, some stars will have up to 13 sectors' worth of data.  A detailed description of the TESS observing strategy can be found in \citet{2015JATIS...1a4003R}. The time series of \citet{2018ApJS..239...34B} reproduce this observing strategy, i.e., their length depends on the position of the respective star in the sky. Here, we shall assume that the TESS mission is extended after its two year nominal duration. We further assume that the observation strategy of the nominal mission will be repeated in possible two year and four year mission extensions, and thus that the same targets are observed again for the same number of sectors after two and four years. However, the scaling relations for p-mode frequency shifts described below are not TESS-specific.
	
	\section{Reconsidering the scaling relation for p-mode frequency shifts}
	There are two models for activity related frequency shifts in the literature: \citet{2007MNRAS.377...17C} simply use the amplitude of the activity cycle $\Delta R'_{\text{HK}}$ (see Equation~\ref{eq:DeltaRHKprime}) as a proxy for the frequency shift and then scale the values such that the solar frequency shift equals the observed $\unit[0.4]{\mu Hz}$. \citet{2002AN....323..357S} found that $\Delta R'_{\text{HK}}$ scales with the average logarithmic fraction of the stellar luminosity that is emitted in the Calcium II H and K spectral lines $R'_{\text{HK}}$: 
	\begin{align}
		\Delta R'_{\text{HK}} \propto {R'_{\text{HK}}}^{0.77}.\label{eq:DeltaRHKprime}
	\end{align}
	For the second model, \citet{2007MNRAS.379L..16M} find that the frequency shifts $\delta\nu$ scale as
	\begin{align}
		\delta\nu \propto \frac{D}{I}\Delta R'_{\text{HK}},
	\end{align}
	where $I$ is the inertia of the modes and $D$ is the depth of the perturbation that causes the mode frequencies to shift. They further assume that the depth $D$ scales as the pressure scale height $H_p$, which is proportional to $L^{0.25} R^{1.5} M^{-1}$ at the stellar photosphere, and that mode inertia scales as $M R^{-1}$, where $R$, $M$, and $L$ are stellar radius, mass, and luminosity respectively. Thus,  \citet{2007MNRAS.379L..16M} find 
	\begin{align}
		\delta\nu \propto \frac{R^{2.5}L^{0.25}}{M^2}\Delta R'_{\text{HK}}\label{eq:Metcalfe_scaling}
	\end{align}
	for the frequency shift amplitude over a full activity cycle.
	
	Here we take a new look at the sensitivity of acoustic mode frequencies to activity. As a starting point we use Equation (1) of \citet{2007MNRAS.379L..16M} to estimate the frequency shift $\delta\nu$:
	\begin{align}
		\delta\nu = \frac{\int dV K S}{2 I \nu},\label{eq:nushift}
	\end{align}
	where $K$ is a kernel for the sensitivity of p modes to perturbations, $S$ is a source function, and $\nu$ is the frequency of the mode. Following \citet{2007MNRAS.379L..16M}, we use
	\begin{align}
		K &\propto \left|\nabla\cdot\vec{\xi}\right|^2,\\
		S &= A\delta(D-D_c),
	\end{align}
	where $\vec{\xi}$ is the eigenfunction of the considered mode, $A$ is the amplitude of the source function, and the $\delta$ function determines the location of the source $D_c$. For now, we set $A=1$ and reintroduce it later via the $R'_{\text{HK}}$ index.
	
	In the following, we concentrate on radial modes, which simplifies the calculations considerably. The eigenfunctions for radial modes can be described by
	\begin{align}
		\vec{\xi} = \xi_r(r) \vec{e}_r, 
	\end{align}
	where $\xi_r(r)$ is the radial displacement eigenfunction and $\vec{e}_r$ is the unit vector in the radial direction. We do not introduce the spherical harmonic $Y_0^0$ into the eigenfunction here, as it would only appear as a constant in the following equations.
	
	With this, the kernel function can be written as
	\begin{align}
		K = \left|\frac{2}{r}\xi_r(r) + \frac{\partial \xi_r(r)}{\partial r}\right|^2.\label{eq:scalekernel}
	\end{align}
	Approximating the derivative $\frac{\partial \xi_r(r)}{\partial r} \approx \frac{\xi_r(r)}{r}$ and assuming that the radial position $r$ which we are considering is close enough to the surface so that the amplitude of the eigenfunction is similar to that at the photospheric stellar radius $R$ we get 
	\begin{align}
		K \approx \frac{9|\xi_r(R)|^2}{R^2}.
	\end{align}
	Thus we have fixed the source to be at $D_c = R$. The mode inertia $I$ can be approximated as follows:
	\begin{align}
		I = \int{dV \rho \left|\vec{\xi}\right|^2}\approx R^3 \overline{\rho} |\xi_r(R)|^2,\label{eq:scaleinertia}
	\end{align}
	where we used again the eigenfunction of a radial mode, replaced the radial integral $\int dr$ by simply multiplying with the photospheric radius $R$, and replaced the density $\rho$ with the mean density $\overline{\rho}$.  
	
	With Equations~(\ref{eq:scalekernel}) and (\ref{eq:scaleinertia}) we can rewrite Equation~(\ref{eq:nushift}):
	\begin{align}
		\delta\nu &\approx \frac{ \frac{9|\xi_r(R)|^2}{R^2}}{2R^3 \overline{\rho} |\xi_r(R)|^2 \nu}\int dV \notag\\
		&=\text{const}\cdot \frac{R}{M\nu}.  \label{eq:modesenspart}
	\end{align}
	Throughout this, we assumed that the source is close enough to the photospheric radius so that $D_c = R$. All constant factors are absorbed into $\text{const}$. Equation~(\ref{eq:modesenspart}) estimates the sensitivity of radial modes to a near-surface perturbation. 
	
	The strength of the perturbation and hence the magnitude of the frequency shift is imparted by the strength of the magnetic activity on the star for which a proxy is used. 
	From the synthetic TESS sample, the $R'_{\text{HK}}$ index can be estimated with a relation found by \citet{1984ApJ...279..763N} for main-sequence stars:
	\begin{align}
		\log\left(\tau_c/P_{\text{rot}}\right) = -\left(0.324 -0.400y +0.283y^2 -1.325y^3\right)\label{eq:noyes},
	\end{align}
	where $y=\log\left(R'_{\text{HK}}\times 10^5\right)$, $\tau_c$ is the convective turnover time, and $P_{\text{rot}}$ is the star's rotation period. Both these values are taken from the models of \citet{2018ApJS..239...34B}. The (local) convective turnover time $\tau_c$ is calculated half a mixing length $\alpha H_p/2$ above the base of the outer convection zone of each star following \citet{1986ApJ...300..339G, 2010A&A...510A..46L}:
	\begin{align}
		\tau_c = \frac{\alpha H_p}{v},
	\end{align}
	where the solar-calibrated mixing length parameter $\alpha=1.957$ is used, $H_p$ is the pressure scale height, and $v$ is the convective velocity. 
	
	To take into account that a large fraction of the stars in the synthetic TESS sample are not on the main-sequence (in contrast to the stars considered in \citet{2002AN....323..357S} and \citet{1984ApJ...279..763N}), we modify the values of  $R'_{\text{HK}}$ with three factors: First, we include a factor  $R^{-2}$ to account for the increasing surface area over which magnetic activity can be distributed, thus affecting the oscillations less. Second, starting at a Rossby number of $\text{Ro}=\frac{P_{\text{rot}}}{\tau_c}=2$, we divide by $\sqrt{\text{Ro}-1}$. This is to accommodate the results of \citet{2017SoPh..292..126M} who found that, as stars reach a critical Rossby number of $\approx 2$, a transition in the magnetic activity appears to happen: Magnetic braking weakens, rotation periods increase more slowly, and cycle periods grow longer and eventually disappear. Finally, we explicitly include the weakening of magnetic activity with the inverse square root of stellar age $t_{\text{Age}}$ $\left[\text{Gyrs}\right]$ found by \citet{Skumanich1972}:
	\begin{align}
		R'_{\text{HK, mod}} = \frac{R'_{\text{HK}}}{ R^2\max\left(\sqrt{\text{Ro}-1},1\right) t_{\text{Age}}^{0.5}}\label{eq:modified_RHK}.
	\end{align}
	
	With this and Equation~(\ref{eq:modesenspart}), the expected frequency shift over a complete stellar activity cycle at the frequency of maximum power $\nu_{\text{max}}$ scales as:
	\begin{align}
		\delta\nu &\propto \frac{R }{M\nu_{\text{max}}}\Delta R'_{\text{HK, mod}}.\label{eq:newscaling}
	\end{align}
	
\begin{figure}
		\begin{center}	\includegraphics[width=0.8\textwidth]{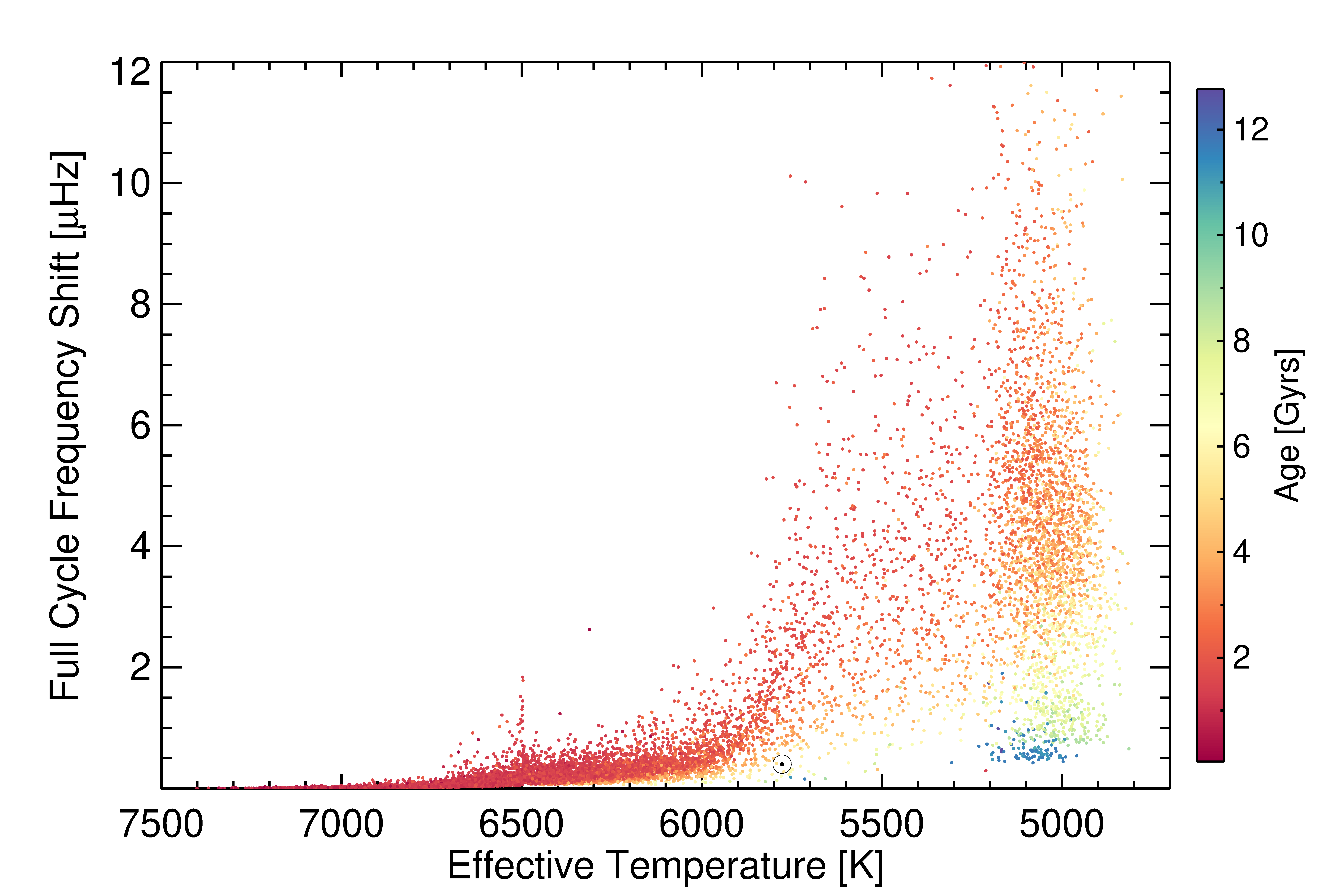}
			\caption{Frequency shift amplitudes for full activity cycles for the synthetic TESS sample calculated with Equation~(\ref{eq:newscaling}). The colour of the dots indicates stellar age. The Sun is indicated by its usual symbol.} 
			\label{fig:cycleshifts}
		\end{center}
	\end{figure}

	Figure~\ref{fig:cycleshifts} shows the expected p-mode frequency shifts for the synthetic TESS sample between the activity minima and maxima of a full stellar activity cycle. The values are scaled to a solar frequency shift of \unit[0.4]{$\mu$Hz} at $\nu_{\text{max}}$. Stellar age is given by the colour of the dots. For the Sun, we assume $R'_{\text{HK}}=-4.901$ (value from \cite{1999ApJ...524..295S}) and do not estimate it with Equation~(\ref{eq:noyes}). The spike in shifts at \unit[6500]{K} is caused by a transition of how rotation rates are modelled by \cite{2018ApJS..239...34B}, cf. their Section 2.5 and their Figure~2.
	
	\section{Measuring $\delta\nu$}\label{ssec:current_techniques}
	Two main methods have been developed to determine the change in p-mode frequency as a function of time. The first technique uses the cross-correlation (CC) between power spectra of different epochs of the time series (see, e.g., \cite{1989A&A...224..253P, 2007ApJ...659.1749C, 2016A&A...589A.103R, 2017A&A...598A..77K}). The regions around the frequency of maximum oscillation amplitude $\nu_{\text{max}}$ of the power spectra are retained and their cross-correlation function is computed. Typically, only the region in which p-mode peaks can be identified is retained. As the expected frequency shifts between two power spectra of any epoch is only of the order of a few $\unit{\mu Hz}$, the complete cross-correlation function does not need to be calculated: restricting the cross-correlation function to a certain lag regime considerably speeds up the computation. The central peak of the cross-correlation function is then fitted to obtain the frequency shift $\delta\nu_{\text{CC}}$ between the two power spectra. To estimate the uncertainty $\sigma$ of the measured frequency shift, we employ the resampling approach described in \cite{2016A&A...589A.103R}.
	
	The second technique uses the individual mode frequencies (peak-bagging: PB) obtained at different epochs. Here, the frequency shift $\delta\nu_{\text{PB}}$, of a particular mode with harmonic degree $\ell$ and radial order $n$ is given by 
	\begin{align}\label{eq:freq_shift}
		\delta\nu_{\text{PB}}=\nu_{\ell,n}(t)-\overline{\nu_{\ell,n}},
	\end{align}
	where $\nu_{\ell,n}(t)$ is the mode frequency obtained from a time series observed between $t$ and $t+\delta t$, and $\overline{\nu_{\ell,n}}$ is a reference frequency for that mode. Both methods (CC and PB) require the data to be split into time series of length $\delta t$, where $\delta t$ is some fraction of the full baseline of observations available. Ideally, $\delta t$ will be sufficiently short to resolve changes with time but long enough that the power spectrum has sufficient resolution and signal-to-noise for the modes to be observed clearly. See \cite{2018ApJS..237...17S} for a detailed description of the peak-bagging procedure in determining frequency shifts.
	
	Most of the targets in the synthetic TESS sample have only few sectors' worth of data (one sector: 7930 stars, two sectors: 2601 stars, three sectors: 781 stars, four sectors or more: 1419; see also Tables~\ref{table:sectors_2yr} and \ref{table:sectors_4yr}). We thus calculate the power spectra for the complete synthetic time series of each star without separating them into shorter segments: here, $\delta t$ is equal to the length of the time series.
	
	In general, the two methods CC and PB are found to produce results in good agreement with each other but the uncertainties associated with the cross-correlation technique tend to be systematically larger than the frequency shifts determined directly from the mode frequencies (see \cite{2007ApJ...659.1749C, 2018ApJS..237...17S} for comparisons of the frequency shift uncertainty obtained with the two methods). However, the peak-bagging method requires good resolution and signal-to-noise for the mode frequencies to be determined accurately and precisely and so there are instances where the cross-correlation method is preferable. The PB technique is also much more reliant on input parameters like initial guesses for mode frequencies and widths than the CC method. We do not apply the PB method here, but assume that the frequency shifts measured with both methods are in agreement and that the uncertainties of the PB method are smaller by a factor of two than those obtained with the CC method.
	
	We implemented the CC technique to measure the frequency shifts in the following way: the power spectrum is calculated from the synthetic time series with an oversampling factor of 8. This power spectrum is then shifted in frequency by the amount that can be expected over the fraction of the stellar cycle that is covered in the two (four) years that pass between the TESS nominal mission and the second observation of any given star (see Section~\ref{ssec:TESS_target}). We estimate the length of the activity cycles by
	\begin{align}
		P_{\text{cyc}} \left[\text{years}\right] = 0.5 P_{\text{rot}} \left[\text{days}\right]\cdot\max\left(\sqrt{\text{Ro}-1}, 1\right)\label{eq:estimate_cycle_period}.
	\end{align}
	We chose this rather crude estimation as a compromise between the short and long activity cycle branches that can be found in a $P_{\text{cyc}}$-$P_{\text{rot}}$ diagram like that of \citet{2007ApJ...657..486B} or \citet{2017SoPh..292..126M}. The lengthening of activity cycles above $\text{Ro}\approx 2$ found by \citet{2017SoPh..292..126M} is again modelled into this with the factor $\max\left(\sqrt{\text{Ro}-1}, 1\right)$. Thus, the input frequency shift is given by
	\begin{align}
		\delta\nu_{\text{input shift}} = \delta\nu \cdot \min\left(\frac{2n}{P_{\text{cyc}}},1\right),
	\end{align}
	where the factor 2 accounts for the fact that the full cycle frequency shift is observed over a half-cycle, between activity minimum and maximum, $n=2$ or 4 is the number of years between observations, depending on the length of the TESS mission extension, and $ \delta\nu$ is calculated with Equation~(\ref{eq:newscaling}). After shifting the power spectrum, we return to only the natural frequencies without oversampling. The oversampling was only introduced to be able to shift by frequencies smaller than the natural frequency resolution, which is $\approx \unit[0.422]{\mu Hz}$ for stars with only one sector of data.
	
	The regions around the frequency of maximum oscillation amplitude $\nu_{\text{max}}$ of the unshifted and shifted power spectra are then cross-correlated. We retain a width of $\pm 5 \Delta\nu$ around $\nu_{\text{max}}$, where $\Delta\nu$ is the large frequency separation. Both $\Delta\nu$ and $\nu_{\text{max}}$ are given in the catalogue of \citet{2018ApJS..239...34B}. To obtain the frequency shift between the two power spectra, we fit the cross-correlation function with a Lorentzian profile.
	
	\section{Influence of assumptions on the predicted shift amplitudes}\label{sec:approximations}
	There are several assumptions made and approximations applied in the derivation of the scaling relation given by Equation~(\ref{eq:newscaling}). The first assumption is that the magnitude of activity-related frequency shifts should be proportional to the strength of stellar magnetic activity. This is certainly justified by the fact that for the Sun there is a clear positive correlation between the p-mode frequency shifts and proxies for magnetic activity as shown by, e.g., \cite{2015SoPh..290.3095B}. Also, in a follow-up study to \cite{2018ApJS..237...17S}, in which they investigate the dependence of the observed frequency shifts on stellar fundamental parameters, Santos et al. (submit.) find that for a sample of 30 solar-like stars, the amplitude of the frequency shifts decreases with decreasing $R'_{\text{HK}}$.  Hence, the $R'_{\text{HK}}$ activity index, which is available for a relatively large number of stars (see, e.g., \cite{1995ApJ...438..269B} and the database of \cite{2018ApJS..236...19E}) and is employed in the scaling relation of \cite{2007ApJ...659.1749C}, is a good starting point for stars similar to the Sun. In principle, spectroscopic activity indices other than $R'_{\text{HK}}$ can be used in any of the frequency shift scaling relations we discuss here; \cite{2007MNRAS.379L..16M}, for example, use the Mg II emission index $i_{\text{Mg II}}$. However, the scaling of the amplitude of the index over a complete activity cycle, as is done for $R'_{\text{HK}}$ in Equation~(\ref{eq:DeltaRHKprime}), has to be adjusted accordingly.
	
	To account not only for the strength of the activity, as is done in the \citeauthor{2007ApJ...659.1749C} scaling, but also for the physics of the oscillation modes, \cite{2007MNRAS.379L..16M} derived the sensitivity of mode frequencies to near surface perturbations. We note that the sensitivity of modes we find with Equation~(\ref{eq:modesenspart}) is in fact very similar to that found by \cite{2007MNRAS.379L..16M} and quoted by \cite{2009MNRAS.399..914K} in their Equation (14). Inserting the scaling relation of \cite{1995A&A...293...87K} $	\nu_{\text{max}} \propto M R^{-1} T_{\text{eff}}^{-0.5}$ into Equation~(\ref{eq:modesenspart}) and using the Stefan-Boltzmann law $\sqrt{T_{\text{eff}}}\propto \left(R^{-2}L\right)^{1/8}$, gives $\delta\nu \approx \frac{R^{2.75} L^{0.125}}{M^2}$. All quantities are assumed to be normalised to solar values in this discussion. The difference of our Equation~(\ref{eq:modesenspart}) to the mode sensitivity factor of \cite{2007MNRAS.379L..16M} is thus only a factor $R^{-0.25} L^{0.125}\propto\sqrt{T_{\text{eff}}}$, which is in the range 0.91--1.14 for the sample of stars investigated here, i.e., is of order unity. Thus, the approximations we made in the derivation of Equation~(\ref{eq:modesenspart}) --- approximation of the radial integral by simply multiplying with the stellar radius $R$, approximation of the derivatives, localising the perturbation to the surface --- yield essentially the same result as the calculations of \cite{2007MNRAS.379L..16M}.
	
	The three scaling relations are most easily distinguished for stars reaching the lower red giant branch, which is where the disagreement is largest. In Figures~\ref{fig:Chaplin} and \ref{fig:Metcalfe} we show the full-cycle frequency shifts of the synthetic TESS sample for the \citeauthor{2007MNRAS.377...17C} and \citeauthor{2007MNRAS.379L..16M} scalings, respectively. The \citet{2007MNRAS.377...17C} relation, simply using the $R'_{\text{HK}}$ estimation of \cite{1984ApJ...279..763N} for all stars from main-sequence to the lower red giant branch, gives maximal full-cycle shift values of $\approx\unit[3]{\mu Hz}$. With estimated cycle lengths of at least 10 years, no shifts larger than $\approx\unit[1.2]{\mu Hz}$ should be found in TESS data with a two year mission extension. This value increases to $\approx\unit[5]{\mu Hz}$ for the new scaling relation presented in this article and to even larger values for the scaling relation of \cite{2007MNRAS.379L..16M}.
	
	As can be seen in Figure~\ref{fig:Metcalfe}, the full-cycle frequency shifts obtained by employing the scaling relation of \cite{2007MNRAS.379L..16M} would lead to very large values of over \unit[50]{$\mu$Hz} for stars reaching the lower red giant branch. This is why we looked for reasonable adjustments. As the sample of stars on which \cite{1984ApJ...279..763N} based their results is exclusively comprised of main-sequence stars, we modified the $R'_{\text{HK}}$ index with the factors discussed above. This aims to capture some of the evolution that can naively be expected in stellar magnetic activity. Here, the decrease of activity with  $t_{\text{Age}}^{-0.5}$ found by \cite{Skumanich1972} is again extrapolated from the main-sequence up to the lower red giant branch. The inclusion of this factor must be understood as an attempt to circumvent our ignorance of how exactly activity evolves after the main-sequence. It is completely possible that activity does indeed decrease faster than just proportional to $t_{\text{Age}}^{-0.5}$ or even stops completely at a certain evolutionary stage. The same applies for the factor $\max\left(\sqrt{\text{Ro}-1}, 1\right)$.
	
	Contemporaneous measurements of the $R'_{\text{HK}}$ index and photometric time series for asteroseismology would facilitate a better differentiation between these scaling relations. If indeed $\delta\nu \propto A \cdot \left(R'_{\text{HK}}\right)^B$, with $A, B$ factors dependent on stellar fundamental parameters regulating mode sensitivity to perturbations ($A$) and the amplitude of activity over a full cycle ($B$), then measurements of frequency shifts and activity level can be used to learn about mode physics as well as stellar dynamos. We note that measurements of stellar rotation periods are essential for the prediction of the $R'_{\text{HK}}$ index with Equation~(\ref{eq:noyes}) or with the modified expression (\ref{eq:modified_RHK}) should direct spectroscopic measurements of stellar activity not be available.
	
	The estimation of the cycle period as described by Equation~(\ref{eq:estimate_cycle_period}) is obviously very crude. If cycle periods are underestimated with this, then the resulting shifts we can expect to observe decrease and our predictions for $\delta\nu$ are an upper limit in this case. More, reliably measured cycle periods are needed to replace this estimation with a more rigorous prediction. We remark however, that Equation~(\ref{eq:estimate_cycle_period}) does include the latest findings concerning the evolution of cycle periods by \cite{2017SoPh..292..126M} through the lengthening of cycles with the factor $\max\left(\sqrt{\text{Ro}-1}, 1\right)$.
	
	\begin{figure}[t]
		\begin{center}
			\includegraphics[width=0.8\textwidth]{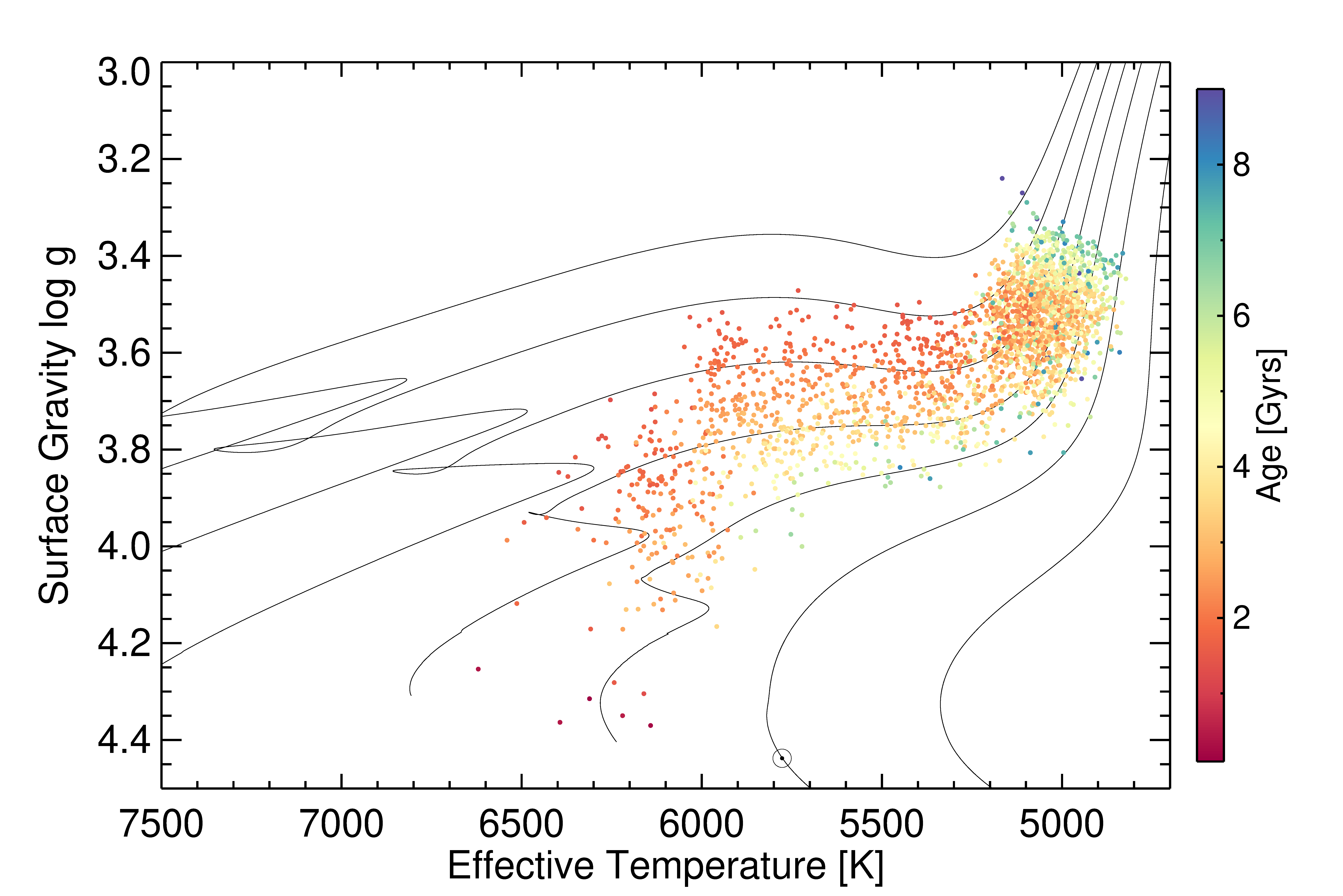}
			\caption{As Figure~\ref{fig:MATL_Kiel} but only stars with significant frequency shifts are shown. Shifts are measured with the cross-correlation technique and a four year TESS mission extension is assumed. Note that the colour bar is different from Figure~\ref{fig:MATL_Kiel}.}
			\label{fig:MATL_Kiel_signif}
		\end{center}
	\end{figure}
	
	\begin{center}
		\begin{table}[t!]
			\caption{Predicted number of stars with significant frequency shifts for different spectral types and evolutionary stages. A four year TESS mission extension is assumed.\\}\label{table:results}
			\begin{tabular}{c|c|c|c|c|c|c}
				\hline
				Spectral type & Region & Number & CC Detections & CC Fraction [\%] & PB Detections & PB Fraction [\%] \\ \hline\hline
				      F       &   I    &  6676  &      141      &       2.1        &      673      &       10.1       \\
				      F       &   II   &  1642  &      24       &       1.5        &      196      &       11.9       \\
				      G       &   I    &   54   &      30       &       55.6       &      44       &       81.5       \\
				      G       &   II   &  1628  &      765      &       47.0       &     1092      &       67.1       \\
				      K       &   II   &  2731  &     1486      &       54.4       &     1799      &       65.9       \\
				     FG       &   I    &  6730  &      171      &       2.5        &      717      &       10.7       \\
				     FGK      &   II   &  6001  &     2275      &       37.9       &     3087      &       51.4       \\ \hline
				    total     &        & 12731  &     2446      &       19.2       &     3804      &       29.9       \\ \hline
			\end{tabular} 
		\end{table}
	\end{center}
	
\section{Potential of TESS}\label{sec:TESS}
We evaluated the frequency shifts obtained from four settings: the TESS mission is either extended by two years or by four years and the frequency shifts are determined by the cross-correlation method (CC) or the peak-bagging method (PB), where the PB shift results are simply taken as those of the CC method but with half the uncertainty as described in Section~\ref{ssec:current_techniques}. The predicted number of detections of significant frequency shifts for different subsets of the cohort of TESS ATL short-cadence stars is given in Table~\ref{table:results} for a four year mission extension and Table~\ref{table:results_2yr} for a two year mission extension.  We define significant frequency shifts as those which are at least $1\sigma>0$, and within $3\sigma$ of the input frequency shift  $\delta\nu_{\text{input shift}}$, where $\sigma$ is the uncertainty of the measured frequency shift. Tightening the second condition to $2\sigma$ or $1\sigma$ would eliminate some of the more precisely but less accurately measured frequency shifts.
	
We separated the spectral types according to their effective temperature with boundaries at $T_{\text{eff}}=\unit[5200]{K}$ between G and K and at $T_{\text{eff}}=\unit[6000]{K}$ between F and G. There are no K-type main-sequence stars in the sample. The separation into region I and II is discussed in Section~\ref{ssec:TESS_target}. Table~\ref{table:results} gives the number of stars for each subset.
	
The suppression of p-mode amplitudes by activity has not been taken into account in our simulations. However, only stars whose oscillations have a low detection probability and small intrinsic mode amplitudes are likely to be affected by this. These are mainly late-type dwarf stars whose number in the mock sample is rather small, thus the overall number of expected detections is not strongly affected by this. As can be seen in Tables~\ref{table:sectors_2yr} and \ref{table:sectors_4yr} the fraction of stars with detectable shifts increases with the length of the time series as frequency resolution increases. However, even for stars which are observed for only one sector, it should be possible to detect shifts at a significant level, especially for stars with long-lived modes.
	
With the CC method we were able to measure significant frequency shifts in 171 stars in region I and in 2275 stars in region II for a four year TESS mission extension. The large difference between the two regions reflects the increased amplitude and longer mode life times of solar-like oscillations for stars in region II. Also, the much smaller mode widths of cooler stars allow one to measure even smaller shifts, especially with the CC method. Figure~\ref{fig:MATL_Kiel_signif} shows a Kiel diagram of the stars with significant shifts measured with the CC method for the four year extension. By comparison of Figure~\ref{fig:MATL_Kiel} with Figure~\ref{fig:MATL_Kiel_signif} it can be seen that stars with detectable frequency shifts are largely found at effective temperatures below $\lesssim$\unit[6500]{K}. Note the different colour tables used in these two figures. For region I, the fraction of F-type stars for which we predict significant shifts is only 2.1\% (141 out of 6676 stars) and only 1.5\% (24 out of 1642) for region II F stars. This reflects the predicted smaller full cycle frequency shifts and large mode widths of F stars. The fraction of stars with predicted detectable frequency shifts increases for G stars: 55.6\% (30 out of 54) for region I and 47.0\% (765 out of 1628) for region II. For the G dwarfs of region I this increase is mainly due to the smaller mode line widths compared to F dwarfs. Furthermore, for region II G-type stars the detectability of shifts is enhanced by the larger predicted full cycle shifts and the larger mode amplitudes. A large fraction of K-type region II stars are predicted to have detectable shifts with 54.4\% (1486 out of 2731). With the smaller uncertainties which we assumed for the PB method, these numbers increase for all subsets of stars, adding up to a total of 3804 out of 12731 stars (29.9\%) for the complete cohort. Table~\ref{table:results_2yr} gives the predicted number of detections for a 2 year TESS extension.

We also tested for the number of false positives caused by the realisation noise. For this we set the input shift to zero for all stars and proceeded as described above. We found no false positives detections with the PB method for either a two or a four year extension even when ommiting the condition that the measured shift be within $3\sigma$ of the input shift value.
	
\begin{figure}
		\begin{center}	
			\includegraphics[width=\textwidth]{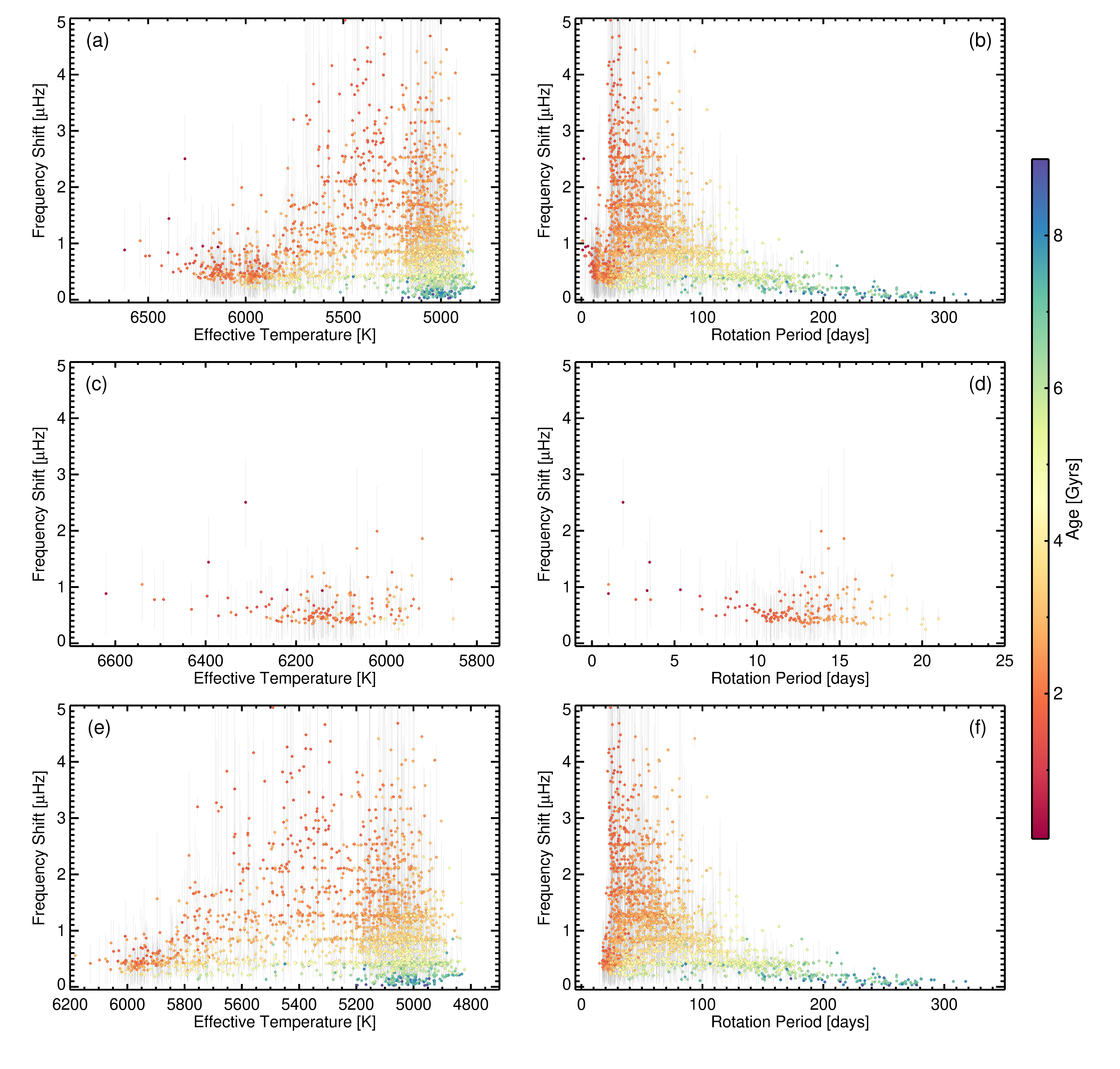}
			\caption{Significant frequency shifts of the synthetic TESS sample for: (a) all stars as a function of effective temperature; (b) all stars as a function of rotation period; (c) stars in region I as a function of effective temperature; (d) stars in region I as a function of rotation period; (e) stars in region II as a function of effective temperature; (f) stars in region II as a function of rotation period. The frequency shifts are measured with the cross-correlation method. Here, a four year TESS mission extension is assumed. The colour of the dots indicates stellar age as given in the colour bar on the right. Note that the range of the abscissa differs from panel to panel.}
			\label{fig:bigplot_age}
		\end{center}
	\end{figure}

\begin{figure}
		\begin{center}	
			\includegraphics[width=\textwidth]{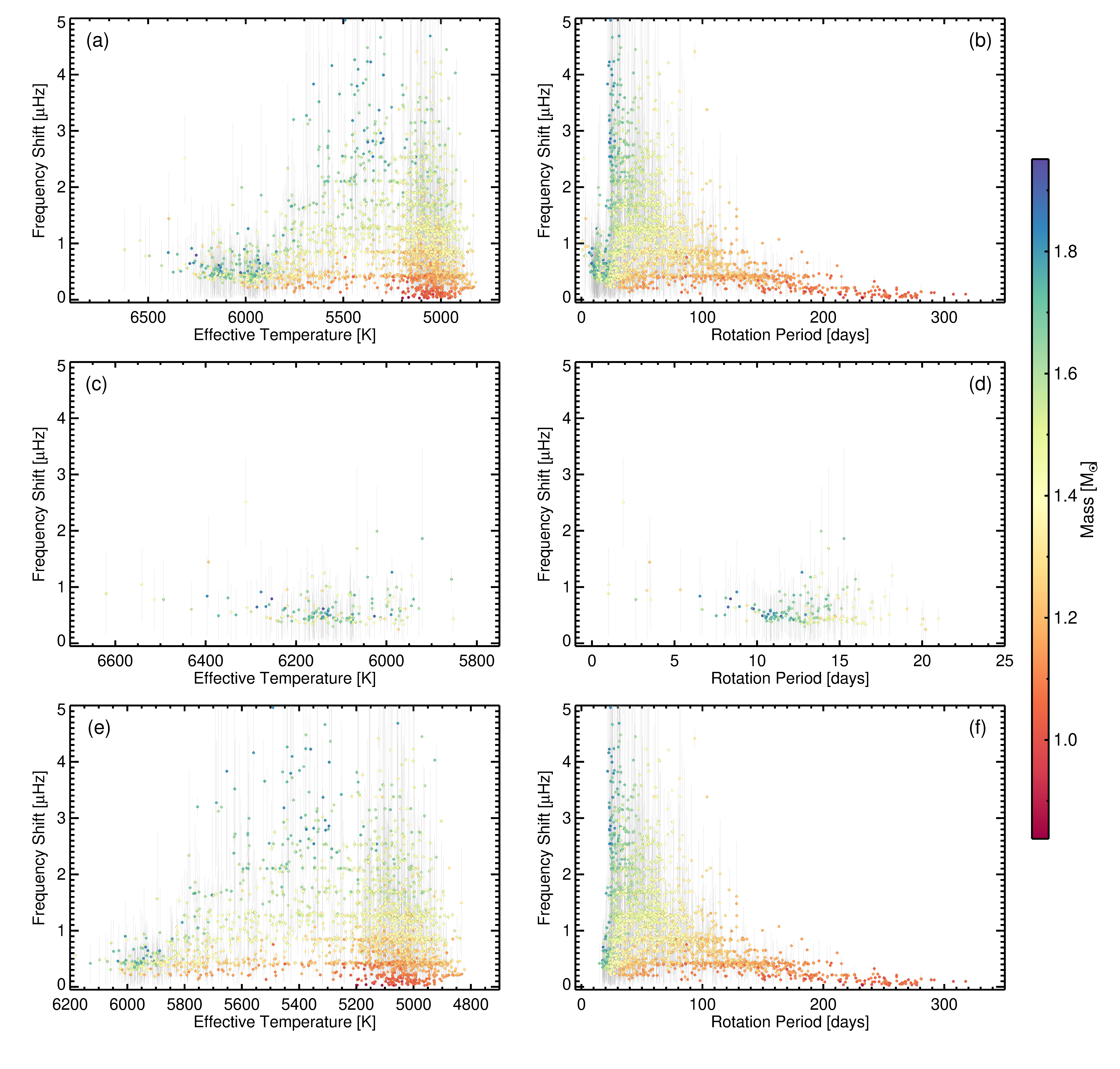}
			\caption{Same as Figure~\ref{fig:bigplot_age} but colour coded with stellar mass.}
			\label{fig:bigplot_mass}
		\end{center}
	\end{figure}

In Figures~\ref{fig:bigplot_age} and \ref{fig:bigplot_mass} the significant frequency shifts obtained with the cross-correlation method for a four year TESS extension are plotted for three subsets of stars (rows from top to bottom: all stars, region I, region II) as a function of effective temperature (left column of panels) and rotation period (right column of panels). The colour of the dots gives stellar age in Figure~\ref{fig:bigplot_age} and stellar mass in Figure~\ref{fig:bigplot_mass}. We capped the ordinate in these plots at $\unit[5]{\mu Hz}$. Higher frequency shifts tend to have large uncertainties and including them would only distort the presentation.
	
By comparing panel (a) of Figure~\ref{fig:bigplot_age} with Figure~\ref{fig:cycleshifts} it can be seen that for older stars, as stellar cycles grow longer with stellar age, only a fraction of the predicted shift over a full activity cycle can be detected with data that are separated by only four years. As expected, younger and faster rotating stars tend to show larger detectable shifts. The older the stars become (green and blue data points), the smaller the shifts that can be expected to be observed after two years become. This can also be seen in panel (b), where the measured shifts are plotted as a function of rotation period: as stars age, their rotation period becomes longer, the level of activity decreases, and the frequency shifts become smaller. Obviously, this by design for our model for the full-cycle frequency shifts (Equation~\ref{eq:newscaling}) with the factors $t_{\text{age}}^{-0.5}$ and the decreasing rotation period with age as it is modelled in the synthetic TESS sample (cf. Section 2.5 and Figure 2 of \cite{2018ApJS..239...34B}). It is noteworthy that for more evolved stars, which have smaller mode damping widths, shifts are detected more precisely and smaller shifts are detectable at a significant level. The lack of stars with a large frequency shift at short rotation periods in panel (b) of Figures~\ref{fig:bigplot_age} and \ref{fig:bigplot_mass} is due to the fact that most of these young, faster-rotating stars in the synthetic TESS sample are hot stars ($T_{\text{eff}}\gtrsim \unit[6500]{K}$) with broader p-mode peaks for which shifts are harder to measure precisely. Also, the full cycle frequency shifts of these stars are rather small as can be seen from Figure~\ref{fig:cycleshifts}.

The discrete structure in the measured frequency shifts in Figures~\ref{fig:bigplot_age} and \ref{fig:bigplot_mass} is due to the large number of stars with only one sector of observation (7930 stars). The frequency resolution of one sector's worth of observation is \unit[0.422]{$\mu$Hz}. As the cross-crorrelation function is computed from periodograms with this resolution, the fitted shift value is likely to be a multiple of this value.
	
For the stars in region I (panels (c) and (d)) there is no clear tendency in frequency shift with effective temperature or rotation period. This is partly due to the way that rotation periods are modelled in the synthetic sample. For stars with $T_{\text{eff}}>\unit[6500]{K}$, most of which are in region I, the rotation period is drawn from a normal distribution with mean \unit[5]{days} and a standard deviation of \unit[2.1]{days}, i.e., there is no dependency on age or effective temperature. Overall, only a small percentage of targets with $T_{\text{eff}}>\unit[6500]{K}$ have predicted detections. The same explanation for this as given above applies. Also, stellar mass introduces a selection bias here: due to their shorter evolutionary time scales, the more massive stars in this sample do not have the time to brake their rotation and thus stars with $M\gtrsim\unit[1.8]{M_{\odot}}$ for which detection of frequency shifts are predicted, are all at rotation periods $P_{\text{rot}}\lesssim\unit[20]{d}$ for stars in region I (panel d) and at $P_{\text{rot}}\lesssim\unit[40]{d}$ for stars in region II (panel f). Similarly, only lower mass stars evolve slowly enough for rotation period to increase and the activity to decrease: note the accumulation of old, low-mass stars at $T_{\text{eff}}\approx \unit[5000]{K}$ in panels (e), which corresponds to the base of the red giant branch seen in Figure~\ref{fig:MATL_Kiel}.\newline

\begin{figure}
	\begin{center}
		\includegraphics[width=0.49\textwidth]{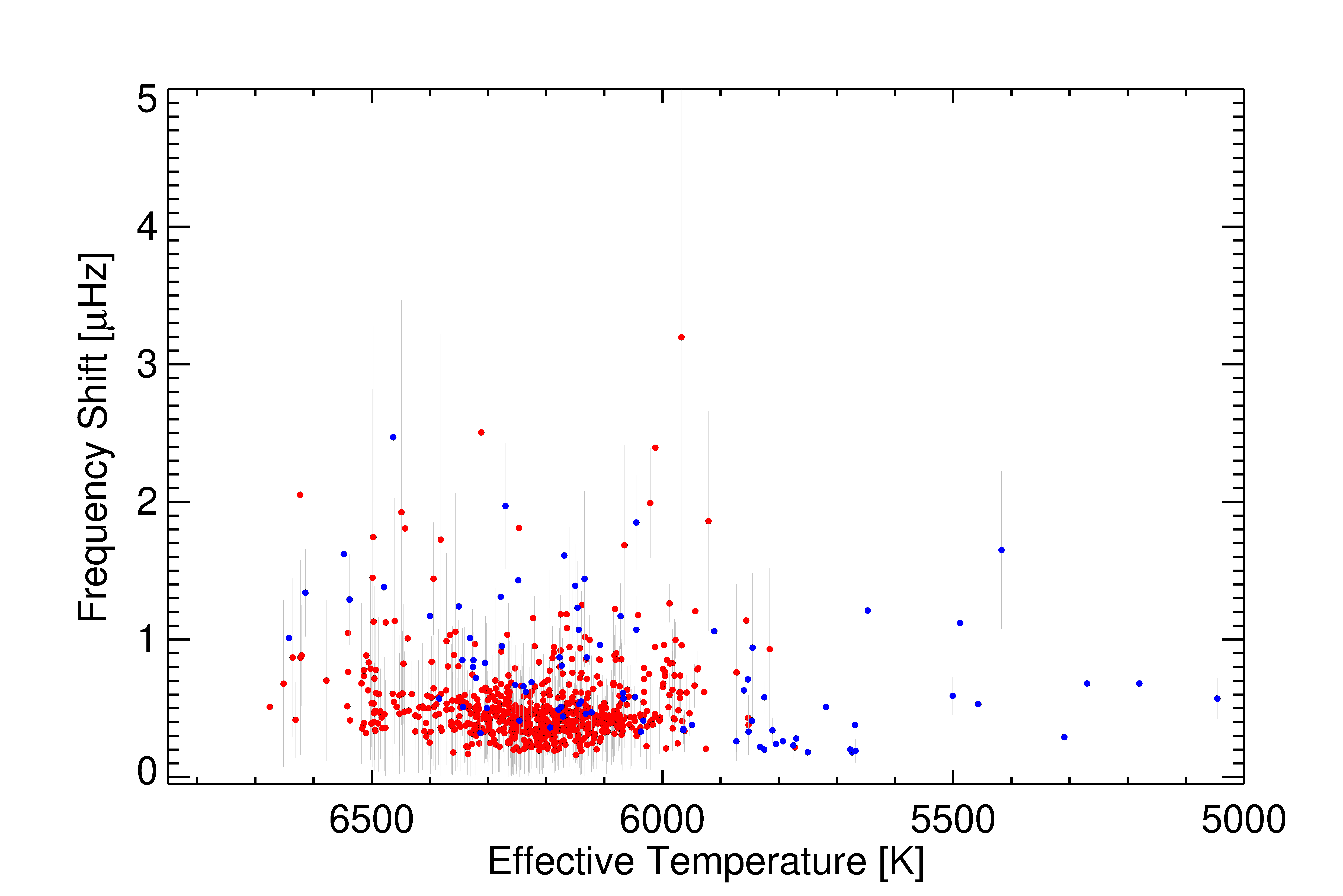}
		\includegraphics[width=0.49\textwidth]{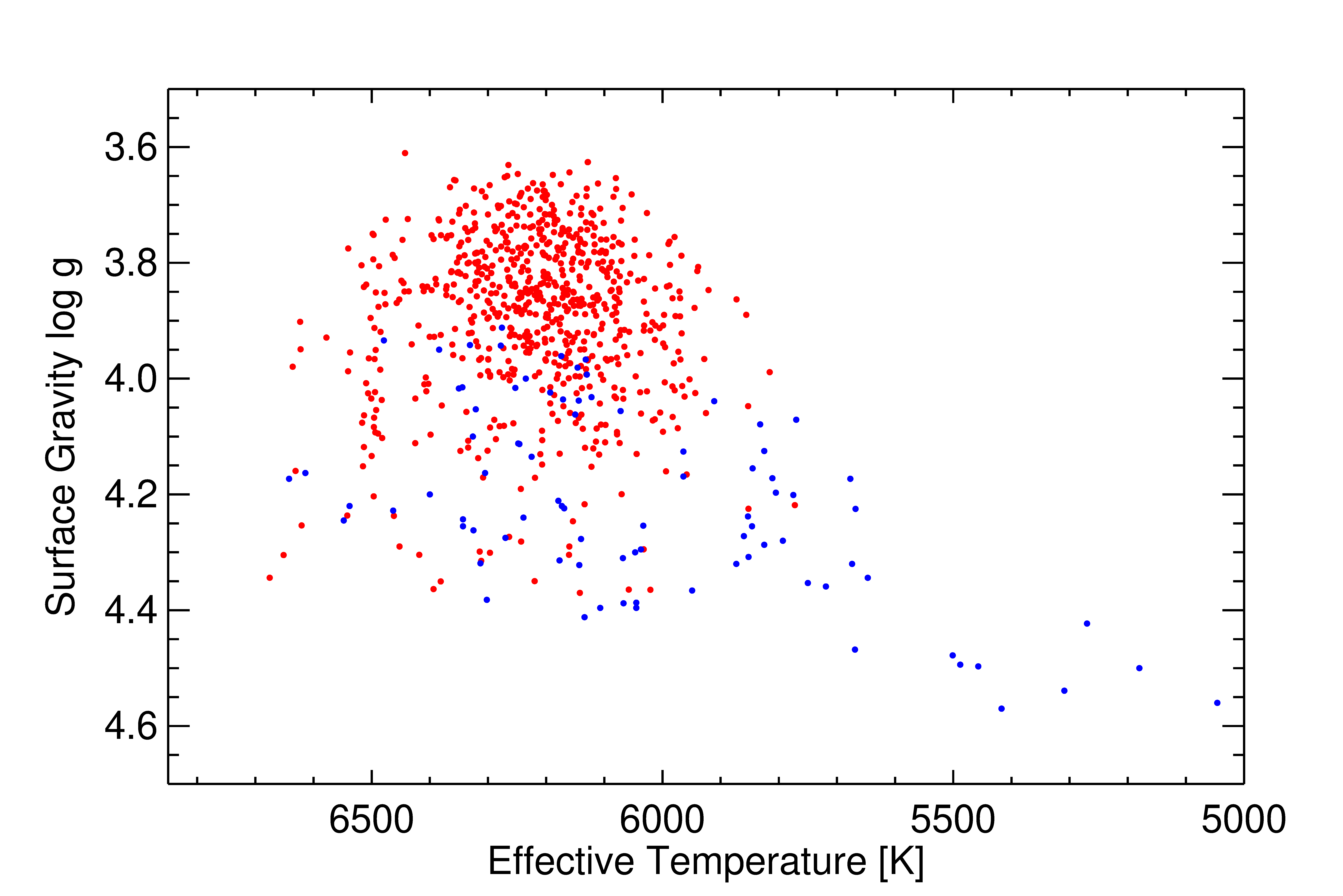}
		\caption{Left panel: Significant frequency shifts from region I of the synthetic TESS data as function of effective temperature in red. A four year extension of the TESS mission is assumed and the error bars of the CC method are halved to approximate the PB method. Significant frequency shifts found by \cite{2018ApJS..237...17S} are shown in blue. Right panel: Kiel diagram of the stars which are included in the left panel.}
		\label{fig:comparison}
	\end{center}
\end{figure}
	
\subsection{Comparison to real observations}
To test the agreement between our predictions and real observations we concentrate on main-sequence stars as there are currently no measured frequency shifts of more evolved stars. We use the results of \cite{2018ApJS..237...17S} available online at this \href{http://vizier.u-strasbg.fr/viz-bin/VizieR-3?-source=J/ApJS/237/17/table2}{link}. This is the largest available record on activity related frequency shifts to date. \cite{2018ApJS..237...17S} employ a PB approach on segments of \textit{Kepler} time series with a length of 90 days which overlap by 45 days. We calculate the frequency shift for each star in their sample by taking the difference between the minimum and maximum value of the mean frequency shift, which combines their results for modes with harmonic degree $l=0$ and 1. 
	
The left panel of Figure~\ref{fig:comparison} shows the shifts of \cite{2018ApJS..237...17S} in blue and the shifts of the region I stars from this work in red as a function of effective temperature. As can be seen, our predictions cover essentially the same range of shifts as the measurements. The baseline of the \textit{Kepler} data used by \cite{2018ApJS..237...17S} is typically around three years, which is shorter than the baseline we assumed for our simulations. Thus, when assuming that the \textit{Kepler} data do not cover a full activity cycle, extrapolating onto the same baseline would give somewhat larger shift values for the measured frequency shifts compared to our predictions. However, our sample is dominated by stars closer to the TAMS as can be seen in the right hand panel of Figure~\ref{fig:comparison}, which shows a Kiel diagram of the stars of \cite{2018ApJS..237...17S} in blue and our region I stars in red. This biases the predicted shifts more towards lower values, as these stars tend to have slowed down their rotation compared to their younger counterparts of similar mass and effective temperature. Restricting our sample further would however leave us with only a very small number of stars. We will have to wait for data of a hopefully extended TESS mission to be able to truly test our scaling relation Equation~(\ref{eq:newscaling}) and those of \cite{2007MNRAS.377...17C} and \cite{2007MNRAS.379L..16M} on a larger sample of stars.

\section{Overestimation of global stellar parameters}\label{sec:overestimate}
The full-cycle frequency shifts as shown in Figure~\ref{fig:cycleshifts} have typical values of $\unit[6]{\mu Hz}$ for a low-luminosity red giant. For such a star, let $\nu_{\text{max}}=\unit[324]{\mu Hz}$, $\Delta\nu=\unit[23.3]{\mu Hz}$, $T_{\text{eff}}=\unit[5058]{K} $, $M=1.064M_{\odot}$ (target 00197 in the sample of \cite{2018ApJS..239...34B}). Using the scaling relation (see, e.g., \citet{1995A&A...293...87K})
	\begin{align}
		M = M_{\odot}\left(\frac{\nu_{\text{max}}}{\nu_{\text{max, }\odot}}\right)^3\left(\frac{\Delta\nu_{\odot}}{\Delta\nu}\right)^4\left(\frac{T_{\text{eff}}}{T_{\text{eff,} \odot}}\right)^{\frac{3}{2}},
	\end{align} 
	the mass value given above is recovered. The solar reference values are those used in \cite{2016ApJ...830..138C}, which are $\nu_{\text{max, }\odot}=\unit[3090]{\mu Hz}$, $\Delta\nu_{\odot}=\unit[135.1]{\mu Hz}$, $T_{\text{eff,} \odot}=\unit[5777]{K}$. If $\nu_{\text{max}}$ is shifted by $\unit[6]{\mu Hz}$ to higher frequencies, a mass of $M_{\text{shift}} = 1.12 M_{\odot}$ is obtained from the scaling relation Thus, if  $\nu_{\text{max}}$ were measured during times of higher activity, stellar mass would be overestimated by $\unit[0.06]{M_{\odot}}$ with this scaling relation. The statistical uncertainties of asteroseismic stellar masses obtained from the standard scaling relations are typically of the order of 10\% \cite{2014ApJS..210....1C, 2016MNRAS.460.4277G}. Thus, an activity-related shift to higher frequencies would introduce a systematic error and lead to an overestimation of mass and radius when using the standard scaling relations. Indeed, \cite{2016ApJ...832..121G} found in their study of 10 eclipsing binary systems with at least one oscillating red giant that the dynamical masses were consistently smaller than those obtained with the asteroseismic scaling relations. The same arguments apply for stellar radii. We note that for more evolved red giants, a shift to higher frequencies of only $\unit[1]{\mu Hz}$ would lead to an even larger overestimation: Considering the values of \cite{2016ApJ...832..121G} for the star KIC~5786154 ($\nu_{\text{max}}=\unit[29.75]{\mu Hz}$, $\Delta\nu=\unit[3.523]{\mu Hz}$, $T_{\text{eff}}=\unit[4747]{K}$) and using the solar reference values and scaling relation as given above, we find a mass for this star of $M=1.44M_{\odot}$. With $\nu_{\text{max}}=\unit[30.75]{\mu Hz}$, which incorporates the frequency shift predicted here, the scaling relation gives $M=1.58M_{\odot}$. This example shows that it might be necessary to calibrate the global seismic parameters $\nu_{\text{max}}$ and $\Delta\nu$ for the level of activity of the star when the scaling relations are to be used for an estimation of mass and radius of more evolved stars. A paper on activity-related frequency shifts in evolved \textit{Kepler} stars is currently in preparation. As we will show there, frequency shifts of a few tenths of \unit{$\mu$Hz} are not uncommon on the red giant branch.
	
	\section{Discussion and Conclusion}\label{sec:discussion}
	To estimate the yield of detections of activity-related p-mode frequency shifts in an extended TESS mission, we first derived a scaling relation for full-cycle shifts (Equation~\ref{eq:newscaling}) for stars from the main-sequence up to low luminosity red giants. We used a catalogue of simulated light curves (\citet{2018ApJS..239...34B}) and input frequency shifts which can be expected to occur within the two (four) years which would pass between a first observation of a given star and a repeated observation by TESS. We then measured these frequency shifts with a cross-correlation technique and found that we can expect to find a couple hundred main-sequence and early subgiant stars and a few thousand more evolved subgiant and low-luminosity red giant stars with activity-related p-mode frequency shifts in an extended TESS mission according to our scaling relation.
	
	Our scaling relation does not account for the varying sensitivity of modes of different harmonic degree: for the Sun it is observed that modes of higher harmonic degree experience larger shifts over the solar cycle \cite[e.g.,][]{2017SoPh..292...67B}. For several \textit{Kepler} stars, larger shifts have been measured for $l=1$ modes than for $l=0$ modes by \cite{2018ApJS..237...17S}.
	A dedicated study of how the frequency shifts of different harmonic degrees can influence the estimation of fundamental stellar parameters is presented in this collection by \citet{2019FrASS...6...41P}.
	
	Up until now there is no confirmed detection of frequency shifts in subgiant and low-luminosity red giant stars. This may either be because their activity cycles are too long to be detectable with the available time series of \textit{Kepler} and CoRoT, they no longer have activity cycles, or nobody has looked for this phenomenon in these stars yet. We also note that it is entirely possible that the scaling relation we have derived here is not valid for stars that have evolved far off the main-sequence even with the adjustments we have included to account for suppression of activity, see Section~\ref{sec:approximations}. 
	
	A dedicated search for activity-related frequency shifts (or frequency offsets in the case of very long cycles, which could also be attributed to activity see Section~\ref{sec:overestimate}), should therefore be carried out with the existing \textit{Kepler} data sets. Such results can then also be used to refine the estimates we presented here. Given our scaling relation, Equation~(\ref{eq:newscaling}) and the large number of confirmed oscillating subgiants and red giants (e.g., \cite{2014ApJS..211....2H, 2018ApJS..236...42Y}) such a study could also help to decide between the scaling relations for the p-mode frequency shifts by \citet{2007MNRAS.377...17C}, \citet{2007MNRAS.379L..16M}, and the one presented here.
	
	So far, only a few dozen solar-like stars, most of them \textit{Kepler} targets, have measured p-mode frequency shifts. An extended sample of seismic detections of stellar activity and activity cycles will also be useful to determine the fundamental parameters which govern stellar dynamos and their evolution as stars age. \cite{2018Sci...361.1231B} have shown that it is possible to measure stellar differential rotation with asteroseismology. This can be connected with photometric measurements of activity and the temporal variation of p-mode frequencies to learn more about the intricate connection of rotation, its change through stellar evolution, and stellar activity cycles. 
	
	Carrying out searches for and investigations of stellar magnetic activity with seismic measures (mode frequencies, and amplitudes) are greatly enhanced in their reliability by contemporaneous spectroscopic measurements of magnetic activity. As, e.g., \cite{2018ApJ...852...46K} have shown, it is feasible to employ seismology together with ground-based observation to study stellar cycles and activity in detail. Stars near the ecliptic poles, which are observed almost continuously during TESS's observation (615 stars in the synthetic sample have 13 sectors of data; a similar number is expected for the real observations) can be used for equivalent studies to fill the $P_{\text{rot}}$-$P_{\text{cyc}}$ diagram. Long-term ground-based spectroscopic measurements of the level of activity of these targets would be expedient for such studies. The rationale to optimally select targets for such campaigns was laid out by \citet{2009MNRAS.399..914K, 2013MNRAS.433.3227K}.  Furthermore, the investigation of stellar variability can strengthen the seismic detection of magnetic activity as, e.g., \cite{2010Sci...329.1032G, 2016A&A...589A.118S} have shown.
	
	Our predictions for frequency shifts measured after four years for stars on the main-sequence and early subgiant stars (region I) agree qualitatively with those found by \cite{2017A&A...598A..77K} and Santos et al. (submit.) for \textit{Kepler} stars of similar evolutionary states. The reason the decrease of shifts with decreasing effective temperature seen by Santos et al. (submit.) is less clear in our simulations is most likely due to a difference in the underlying sample of stars. We will have to await TESS data from a hopefully extended mission, calculate models of the observed stars, measure the frequency shifts, and then compare predictions from our scaling relation and those of \cite{2007MNRAS.377...17C} and \cite{2007MNRAS.379L..16M} to these observed shifts. As also mentioned above, measurements of a spectroscopic activity proxy like $R'_{\text{HK}}$ and reliable measurements of the stellar rotation periods and age would facilitate this considerably.
	
	We found that it should be possible to reliably measure even small average frequency shifts between two TESS observations each with a length of only one month. This should also enable the search for photometric and asteroseismic signatures of short activity cycles in those young, fast rotating stars which have many sectors of data in TESS observations (see \cite{2014MNRAS.441.2744V, 2017A&A...603A..52R} for such studies with \textit{Kepler} data). Obviously, long and uninterrupted photometric time series with high stability are optimal for asteroseismology in general and specifically for the search for seismic signatures of activity cycles. Even though most of the targets of TESS only have few consecutive sectors of observation, time series provided by TESS, especially by an extended mission, can be used for such searches. In the future, PLATO \cite{2014ExA....38..249R} will pick up the torch, observe two target fields for 2-3 years each, and provide data for asteroseismology. These data will make another major contribution in the seismic sounding of stellar activity cycles.

	\section*{Conflict of Interest Statement}
	The authors declare that the research was conducted in the absence of any commercial or financial relationships that could be construed as a potential conflict of interest.
	
	\section*{Author Contributions}
	RK led the work on this article and wrote most of the text. AMB contributed to the writing of Sections~\ref{sec:intro}, \ref{ssec:current_techniques}, and the overall refinement of the writing. WHB generated and provided the dataset which is utilised in this article, computed additional necessary quantities, and helped in the writing of the relevant sections.
	
	\section*{Funding}
	This work was funded by STFC consolidated grant ST/L000733/1.  WHB acknowledges support from the UK Science and Technology Facilities Council (STFC). 
	
	\section*{Acknowledgements}
	RK \& AMB acknowledge the support of the STFC consolidated grant ST/L000733/1.
	
	\section*{Data Availability Statement}
	The datasets used for this study can be found in this repository: \href{https://doi.org/10.5281/zenodo.1470154}{Link}.
	
	\bibliographystyle{frontiersinHLTH_FPHY} 
	\bibliography{references_bib}
	
	\newpage
	
	\appendix
	\section{Additional plots and tables}
	
\begin{figure}[h]
	\begin{center}
		\includegraphics[width=0.8\textwidth]{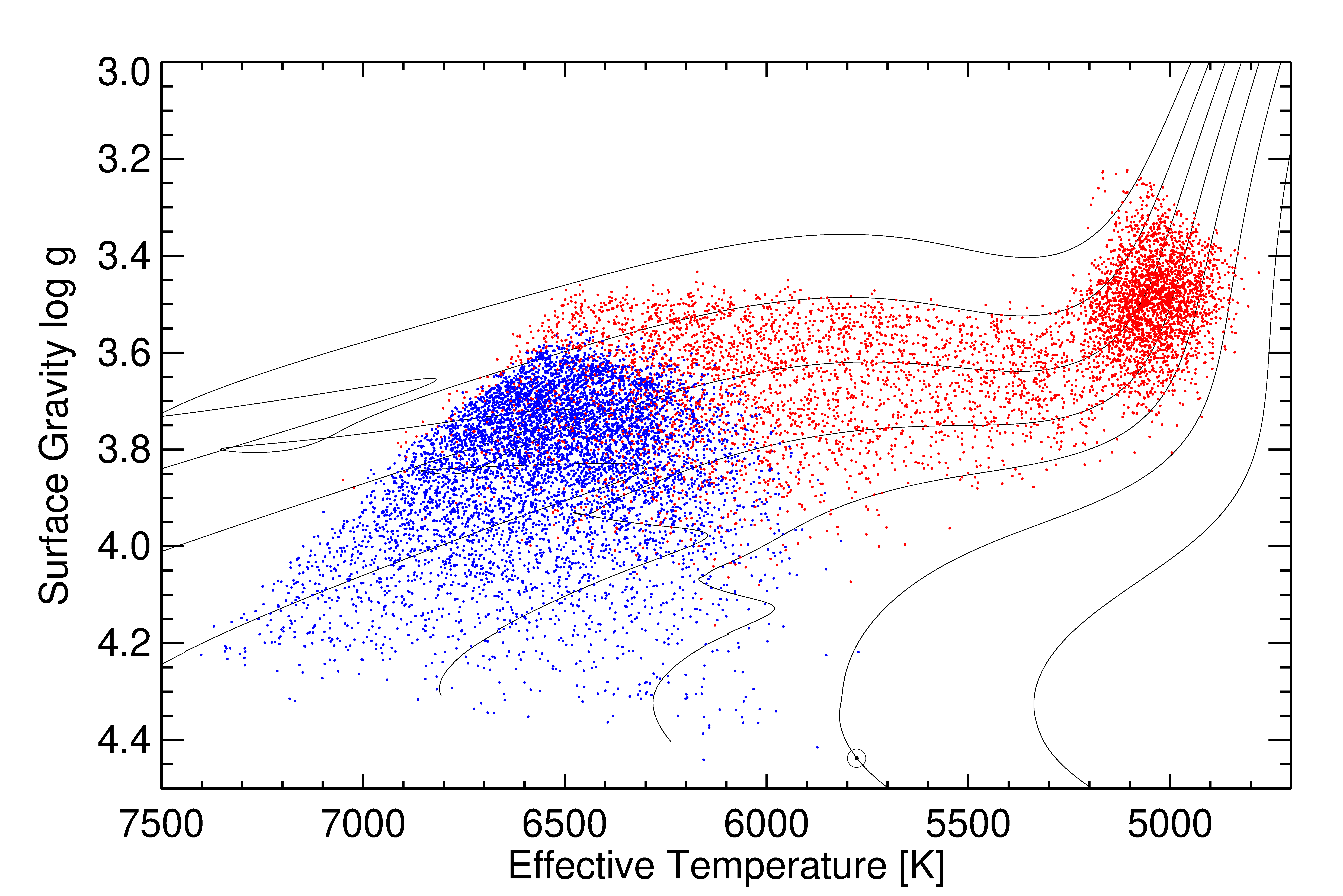}
		\caption{Kiel diagram of the synthetic TESS sample of \citet{2018ApJS..239...34B}. The Sun is indicated by its usual symbol. Evolutionary tracks for stars with masses from 0.8 to \unit[2.0]{M$_{\odot}$} in steps of \unit[0.2]{M$_{\odot}$} are overlaid as solid black lines. Stars on the main-sequence are blue, stars after the Terminal Age Main-Sequence (TAMS) are red. Stars for which the central hydrogen abundance $Y_c$ is below $10^{-5}$ are here defined to have crossed the TAMS.}
		\label{fig:MATL_Kiel_TAMS}
	\end{center}
\end{figure}

\begin{figure}[!h]
		\begin{center}
			\includegraphics[width=0.8\textwidth]{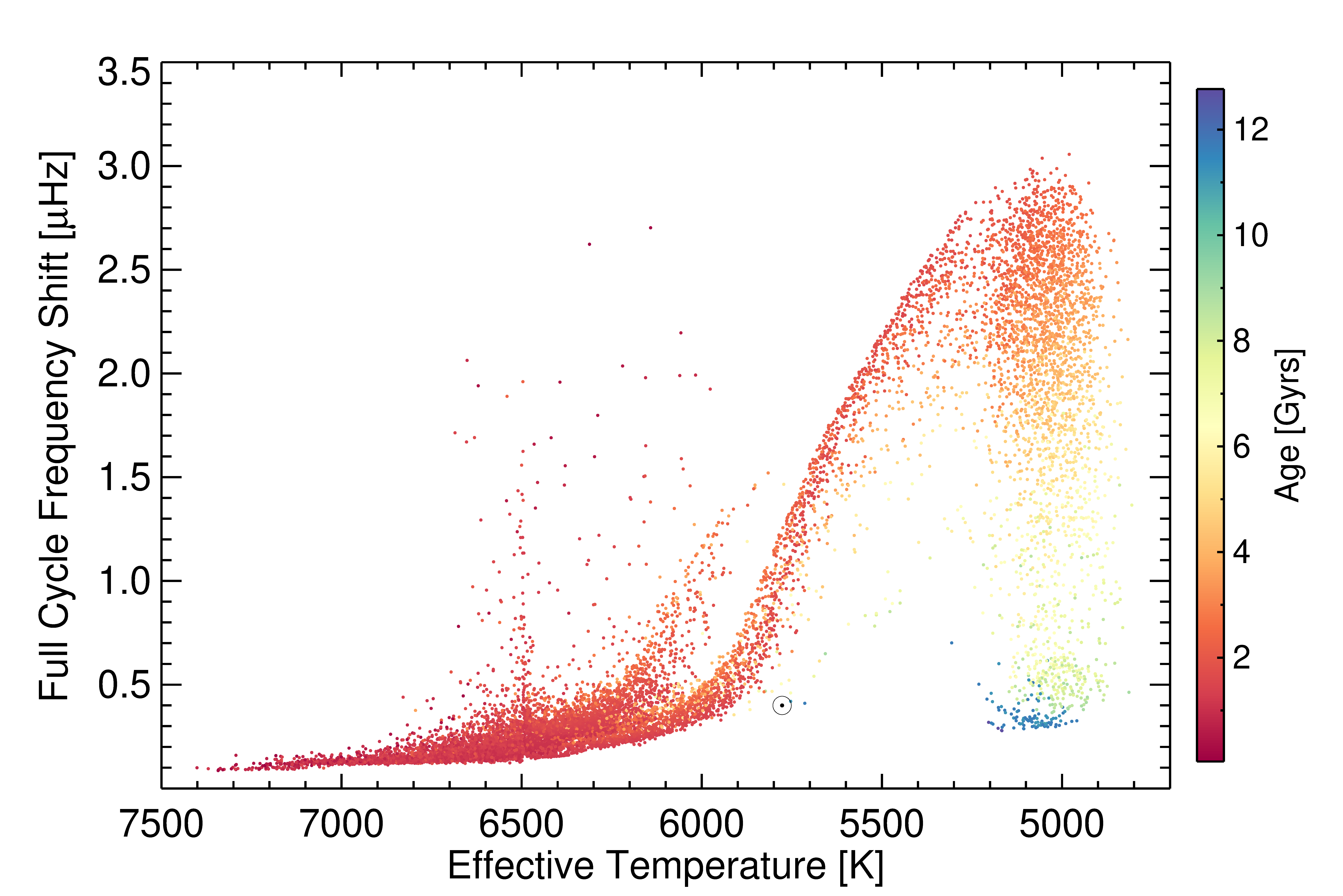}
			\caption{As Figure~\ref{fig:cycleshifts} but for the full-cycle frequency shifts according to the scaling relation of \cite{2007MNRAS.377...17C}.}
			\label{fig:Chaplin}
		\end{center}
\end{figure}
	
\begin{figure}[!h]
		\begin{center}
			\includegraphics[width=0.8\textwidth]{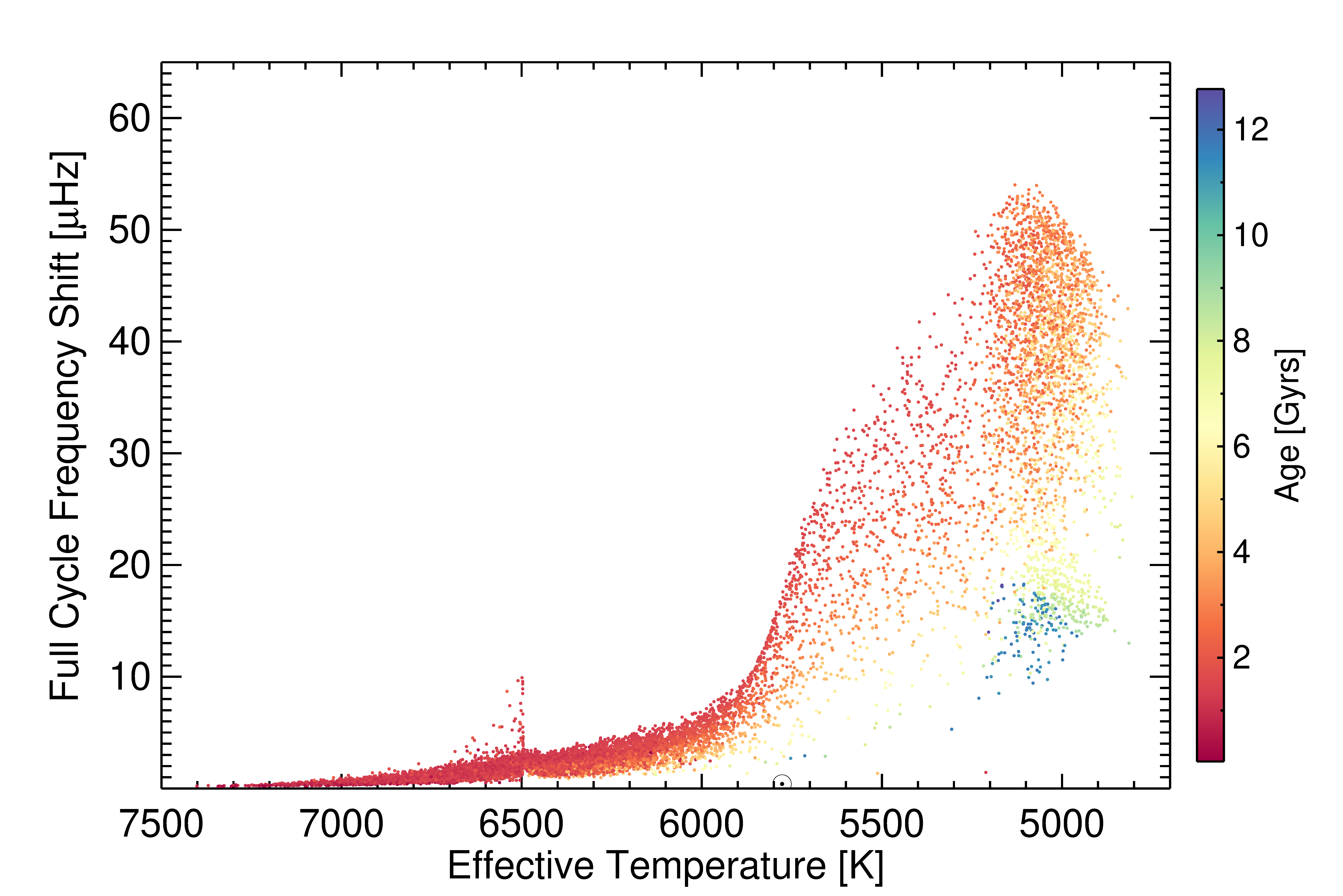}
			\caption{As Figure~\ref{fig:MATL_Kiel} but for the full-cycle frequency shifts according to the scaling relation of \cite{2007MNRAS.379L..16M} given in Equation~(\ref{eq:Metcalfe_scaling}).}
			\label{fig:Metcalfe}
		\end{center}
\end{figure}
	
\begin{sidewaystable}
	\begin{center}
			\caption{Number of stars with $N$ consecutive sectors of data, number of significant shift detections, and fraction of detections of the total number of stars, for all stars in the sample, the main-sequence stars (region I), and post-main-sequence stars (region II). Shifts measured with the CC and PB methods for a two year TESS mission extension.\\}\label{table:sectors_2yr}
		\begin{tabular}{c|c|c|c|c|c|c|c|c|c|c|c|c|c}
			\hline
			       Sectors  $N$         &  1   &  2   &  3   &  4   &  5   &  6  &  7  &  8   &  9   &  10  &  11  &  12  &  13  \\ \hline\hline
			       Total number         & 7930 & 2601 & 781  & 306  &  9   &  0  &  0  &  40  & 144  & 105  & 114  &  86  & 615  \\
			      CC   Detections       & 1101 & 595  & 232  &  98  &  3   &  0  &  0  &  12  &  53  &  38  &  44  &  32  & 141  \\
			     CC Fraction [\%]       & 13.9 & 22.9 & 29.7 & 32.0 & 33.3 & 0.0 & 0.0 & 30.0 & 36.8 & 36.2 & 38.6 & 37.2 & 22.9 \\
			      PB  Detections        & 1749 & 782  & 284  & 114  &  4   &  0  &  0  &  14  &  56  &  45  &  48  &  37  & 172  \\
			     PB Fraction [\%]       & 22.1 & 30.1 & 36.4 & 37.3 & 44.4 & 0.0 & 0.0 & 35.0 & 38.9 & 42.9 & 42.1 & 43.0 & 28.0 \\ \hline
			         Region I           & 4251 & 1391 & 393  & 155  &  3   &  0  &  0  &  20  &  69  &  53  &  50  &  42  & 303  \\
			  CC  Detections region I   &  57  &  17  &  8   &  2   &  0   &  0  &  0  &  0   &  2   &  3   &  0   &  1   &  4   \\
			CC Fraction region I   [\%] & 1.3  & 1.2  & 2.0  & 1.3  & 0.0  & 0.0 & 0.0 & 0.0  & 2.9  & 5.7  & 0.0  & 2.4  & 1.3  \\
			  PB  Detections region I   & 287  & 104  &  30  &  8   &  0   &  0  &  0  &  1   &  2   &  7   &  2   &  3   &  14  \\
			PB Fraction region I   [\%] & 6.8  & 7.5  & 7.6  & 5.2  & 0.0  & 0.0 & 0.0 & 5.0  & 2.9  & 13.2 & 4.0  & 7.1  & 4.6  \\ \hline
			         Region II          & 3679 & 1210 & 388  & 151  &  6   &  0  &  0  &  20  &  75  &  52  &  64  &  44  & 312  \\
			 CC  Detections region II   & 1044 & 578  & 224  &  96  &  3   &  0  &  0  &  12  &  51  &  35  &  44  &  31  & 137  \\
			CC Fraction region II [\%]  & 28.4 & 47.8 & 57.7 & 63.6 & 50.0 & 0.0 & 0.0 & 60.0 & 68.0 & 67.3 & 68.8 & 70.5 & 43.9 \\
			 PB  Detections region II   & 1462 & 678  & 254  & 106  &  4   &  0  &  0  &  13  &  54  &  38  &  46  &  34  & 158  \\
			PB Fraction region II [\%]  & 39.7 & 56.0 & 65.5 & 70.2 & 66.7 & 0.0 & 0.0 & 65.0 & 72.0 & 73.1 & 71.9 & 77.3 & 50.6 \\ \hline
		\end{tabular} 
	\end{center}
\end{sidewaystable}

\begin{sidewaystable}[h]
	\begin{center}
		\caption{Number of stars with $N$ consecutive sectors of data, number of significant shift detections, and fraction of detections of the total number of stars, for all stars in the sample, the main-sequence stars (region I), and post-main-sequence stars (region II). Shifts measured with the CC and PB methods for a four year TESS mission extension.\\}\label{table:sectors_4yr}
	\begin{tabular}{c|c|c|c|c|c|c|c|c|c|c|c|c|c}
		\hline
		       Sectors  $N$         &  1   &  2   &  3   &  4   &  5   &  6  &  7  &  8   &  9   &  10  &  11  &  12  &  13  \\ \hline\hline
		       Total number         & 7930 & 2601 & 781  & 306  &  9   &  0  &  0  &  40  & 144  & 105  & 114  &  86  & 615  \\
		      CC   Detections       & 1139 & 628  & 248  & 102  &  3   &  0  &  0  &  12  &  55  &  39  &  39  &  33  & 148  \\
		     CC Fraction [\%]       & 14.4 & 24.1 & 31.8 & 33.3 & 33.3 & 0.0 & 0.0 & 30.0 & 38.2 & 37.1 & 34.2 & 38.4 & 24.1 \\
		      PB  Detections        & 2094 & 874  & 311  & 126  &  4   &  0  &  0  &  14  &  65  &  45  &  47  &  40  & 184  \\
		     PB Fraction [\%]       & 26.4 & 33.6 & 39.8 & 41.2 & 44.4 & 0.0 & 0.0 & 35.0 & 45.1 & 42.9 & 41.2 & 46.5 & 29.9 \\ \hline
		         Region I           & 4251 & 1391 & 393  & 155  &  3   &  0  &  0  &  20  &  69  &  53  &  50  &  42  & 303  \\
		  CC  Detections region I   & 108  &  30  &  14  &  6   &  0   &  0  &  0  &  0   &  0   &  3   &  1   &  3   &  6   \\
		CC Fraction region I   [\%] & 2.5  & 2.2  & 3.6  & 3.9  & 0.0  & 0.0 & 0.0 & 0.0  & 0.0  & 5.7  & 2.0  & 7.1  & 2.0  \\
		  PB  Detections region I   & 473  & 140  &  42  &  20  &  0   &  0  &  0  &  1   &  5   &  7   &  2   &  4   &  23  \\
		PB Fraction region I   [\%] & 11.1 & 10.1 & 10.7 & 12.9 & 0.0  & 0.0 & 0.0 & 5.0  & 7.2  & 13.2 & 4.0  & 9.5  & 7.6  \\ \hline
		         Region II          & 3679 & 1210 & 388  & 151  &  6   &  0  &  0  &  20  &  75  &  52  &  64  &  44  & 312  \\
		 CC  Detections region II   & 1031 & 598  & 234  &  96  &  3   &  0  &  0  &  12  &  55  &  36  &  38  &  30  & 142  \\
		CC Fraction region II [\%]  & 28.0 & 49.4 & 60.3 & 63.6 & 50.0 & 0.0 & 0.0 & 60.0 & 73.3 & 69.2 & 59.4 & 68.2 & 45.5 \\
		 PB  Detections region II   & 1621 & 734  & 269  & 106  &  4   &  0  &  0  &  13  &  60  &  38  &  45  &  36  & 161  \\
		PB Fraction region II [\%]  & 44.1 & 60.7 & 69.3 & 70.2 & 66.7 & 0.0 & 0.0 & 65.0 & 80.0 & 73.1 & 70.3 & 81.8 & 51.6 \\ \hline
	\end{tabular} 
	\end{center}
\end{sidewaystable}

	\begin{center}
		\begin{table}[h]
			\caption{Detectability of frequency shifts for different samples of stars. A two year TESS mission extension is assumed.\\}\label{table:results_2yr}
			\begin{tabular}{c|c|c|c|c|c|c}
				\hline
				Spectral type & Region & Number & CC Detections & CC Fraction [\%] & PB Detections & PB Fraction [\%] \\ \hline\hline
				      F       &   I    &  6676  &      62       &       0.9        &      416      &       6.2        \\
				      F       &   II   &  1642  &       6       &       0.4        &      49       &       3.0        \\
				      G       &   I    &   54   &      32       &       59.3       &      42       &       77.8       \\
				      G       &   II   &  1628  &      775      &       47.6       &     1120      &       68.8       \\
				      K       &   II   &  2731  &     1474      &       54.0       &     1678      &       61.4       \\
				     FG       &   I    &  6730  &      94       &       1.4        &      458      &       6.8        \\
				     FGK      &   II   &  6001  &     2255      &       37.6       &     2847      &       47.4       \\ \hline
				    total     &        & 12731  &     2349      &       18.5       &     3305      &       26.0       \\ \hline
			\end{tabular} 
		\end{table}
	\end{center}
	
	
\end{document}